\documentclass{article}

\usepackage{arxiv}

\usepackage[utf8]{inputenc} % allow utf-8 input
\usepackage[T1]{fontenc}    % use 8-bit T1 fonts
\usepackage{hyperref}       % hyperlinks
\usepackage{url}            % simple URL typesetting
\usepackage{booktabs}       % professional-quality tables
\usepackage{amsfonts}       % blackboard math symbols
\usepackage{nicefrac}       % compact symbols for 1/2, etc.
\usepackage{microtype}      % microtypography
\usepackage{lipsum}
\usepackage{graphicx}

\usepackage{algorithmic}
\usepackage{subcaption}
\usepackage{textcomp}
\usepackage{xcolor}
\usepackage{caption}
\usepackage{setspace}
\usepackage{mathptm}
\usepackage{amsmath}
\usepackage{booktabs}
\usepackage{multirow}

\usepackage[ruled,vlined]{algorithm2e}
\usepackage{circuitikz}
\usepackage{tikz}
\usetikzlibrary{shapes,arrows,shadows,decorations.pathreplacing,backgrounds,calc}

\title{ProbLock: Probability-based Logic Locking}

\author{
  Michael~Yue \\
  Department of Electrical and Computer Engineering\\
  Santa Clara University\\
  Santa Clara, California, USA \\
  \texttt{myue@scu.edu} \\
   \And
 Fatemeh~Tehranipoor \\
  Department of Electrical and Computer Engineering\\
  Santa Clara University\\
  Santa Clara, California, USA \\
  \texttt{ftehranipoor@scu.edu} \\

}

\begin{document}
\maketitle

\begin{abstract}
Integrated circuit (IC) piracy and overproduction are serious issues that threaten the security and integrity of a system. Logic locking is a type of hardware obfuscation technique where additional key gates are inserted into the circuit. Only the correct key can unlock the functionality of that circuit otherwise the system produces the wrong output. In an effort to hinder these threats on ICs, we have developed a probability-based logic locking technique to protect the design of a circuit. Our proposed technique called ``ProbLock'' can be applied to combinational and sequential circuits through a critical selection process. We used a filtering process to select the best location of key gates based on various constraints. Each step in the filtering process generates a subset of nodes for each constraint. We also analyzed the correlation between each constraint and adjusted the strength of the constraints before inserting key gates. We have tested our algorithm on 40 benchmarks from the ISCAS '85 and ISCAS '89 suite.
\end{abstract}

% keywords can be removed

\section{Introduction and Background}
\begin{figure}
    \centering
    \begin{subfigure}[b]{0.4\columnwidth}
        \centering
        \includegraphics[scale=.5]{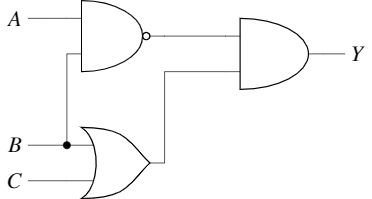}
        \caption{Unlocked Circuit}
        \label{fig:logic_locking_1}
    \end{subfigure}
    \hfill %%
    \begin{subfigure}[b]{0.475\columnwidth}
        \centering
        \includegraphics[scale=.48]{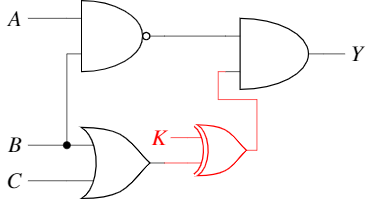}
        \caption{locked Circuit}
        \label{fig:logic_locking_2}
    \end{subfigure}
    \caption{An example of logic locking circuit.}
    \label{fig:logic_locking}
\end{figure}

The semiconductor industry is constantly changing from the production of ICs to the complexity of their design. The industry has moved to a fabless model where most of the fabrication for a chip is outsourced a less secure and less trusted environment. These environments include testing and fabrication facilities that are necessary for the pipeline. While this model does improve production costs and development, it has also led to the consequence of piracy, overproduction, and cloning. The chips are also vulnerable to various attacks~\cite{tehranipoor2018low} that attempt to extract the design of the chip or other information from the device. Due to these security issues, researchers have developed techniques to counter these attacks.  One technique to improve the security of ICs is hardware obfuscation~\cite{book}. Hardware obfuscation is a technique that modifies the structure or description of a circuit in order to make it harder for an attacker to reverse engineer the hardware. Some obfuscation techniques modify the gate level structure of the circuit while other techniques add gates to protect the logic of the circuit. Logic locking is a technique that inserts additional gates and logic components into a circuit which will lock the circuit and produce an incorrect output unless the proper key is provided to the circuit. The IC will be considered locked or functionally incorrect until the correct key unlocks the additional gates. Using XOR and XNOR components as key gates, the proper key value will make the gate act as a buffer and have no effect on the rest of the logic. If the wrong key value is provided, the key gate will produce a wrong value and make the circuit nonfunctional. Figure~\ref{fig:logic_locking} shows an example of logic locking. A key gate is added in between logic gates with one input connected to the key bit value. The addition of these key gates adds a small overhead to the overall circuit while increasing the security of the device.

Various techniques have been proposed by other researchers to protect the integrity and privacy of integrated circuits. Logic cone analysis was used to develop a logic locking technique in 2015~\cite{logiccone}. This technique used fan-in and fan-out metrics to insert key gates into a netlist. Logic cone analysis was vulnerable to SAT attacks that were developed over the next few years. SAT attacks probed the input and output patterns of the system to determine the key to unlock a circuit. SAT attacks have proven to be very effective against logic locking techniques. Strong logic locking (SLL) was another technique developed that analyzed the relationship between inserted key gates in the form of a graph~\cite{onimproving}~\cite{overview}. SLL was also vulnerable to SAT-based attacks. A new SAT resistant technique called SARLock was later implemented with the main purpose of thwarting SAT attacks~\cite{sarlock}. The method made SAT attacks exponential in complexity and therefore ineffective. Tehranipoor et al.~\cite{tehranipoor2019deep} explores the potential of employing the state-of-the-art deep RNN that allows an attacker to derive the correct values of the key inputs (secret key) from logic encryption hardware obfuscation techniques.

All of the logic locking discussed is effective in improving security for an IC, however, they are still vulnerable to sensitization exploits and strong oracle based attacks. 

To overcome the aforementioned issues, in this paper we propose a very new logic locking technique which we call \textbf{"ProbLock"}. ProbLock is a probability-based technique that inserts key gates into a circuit netlist where only the correct key value will unlock the circuit. In this paper, we propose this technique as a form of logic locking where each step of the process narrows down a set of best nodes to insert key gates. We used four constraints to filter out the best nodes and choose the location of the inserted key gates. A probability constraint is the main metric that we used to lock the circuits. We tested our technique by obfuscating a set of circuit benchmarks from ISCAS '85 and ISCAS '89 suite~\cite{iscas85}~\cite{iscas89}. These include a variety of combinations and sequential circuits. We analyzed the relationship and correlation between constraints in our technique and found some relationships that support the strength of our technique. 

Specifically, we have the following contributions in this paper:
\begin{itemize}
    \item We present a probability-based logic locking (ProbLock) technique to lock a circuit with low overhead using a filtering process. 
    \item We implemented a design where the strength of the filtering process can be adjusted for different situations.
    \item We analyzed the correlation between constraints and showed how the relationship between constraints can strengthen the security process. 
    \item We obfuscated 40 benchmarks from ISCAS '85 and ISCAS '89 using ProbLock.
\end{itemize}

\section{Literature Review}
Many techniques of logic locking have already been proposed and tested against certain attacks and on circuit benchmarks. One of the earliest logic locking techniques inserted key gates randomly into the circuit. This provided some security, but many attacks were developed to break this method. Another obfuscation technique was developed using logic cone analysis in~\cite{logiccone}. Sections of a circuit can be grouped into logic cones by calculating the fan-in and fan-out values of a gate. Inserting key gates at certain logic cones areas will increase the security of the system. Logic cone analysis is good for countering logic cone attacks. Certain attacks will exploit these weak logic cones and try to discover the key to unlock the circuit. Logic cone analysis is vulnerable to other types of attacks such as SAT and functional attacks. 
 
 %\noindent{\textbf{SAT Attack}:} Boolean satisfiability or SAT attack is a type of algorithm that intends to determine the key based on certain inputs and corresponding outputs to the circuit. A distinguishing input patterns (DIP) is the metric used to determine the viability of a SAT attack. DIPs will eliminate wrong key values to unlock the circuit. Different input output patterns reveal potential key bit values for the locked circuit. SAT attacks are strong against obfuscation techniques that are not designed specifically to counter it. 
 
 Strong logic locking (SLL) is another obfuscation technique, but it is also vulnerable to SAT attacks~\cite{onimproving}. SLL is based on interference graphs that show how inserted key gates interfere with each other. The interference graph shows the relationship between an inserted key gate and its surrounding key gates and wires. The interference graph shows if key gates are on a cascading path, parallel path, or if they don't interfere with each other at all. The interference graph along with other information makes it harder for an attacker to unlock the circuit even with a SAT attack model.

More recent techniques have been developed to counter SAT attacks and other related schemes. The obfuscation technique needs to be strong enough to resist certain attacks otherwise the integrity of the IC would be compromised. The goal of an adversary during an attack is to determine the secret key to unlock the circuit or gain other important information from the system. SARLock was developed to make the SAT attack model inefficient~\cite{sarlock}. SARLock employs a small overhead strategy that exponentially increases the number of DIPs needed to unlock the circuit. SARLock is very strong against SAT attacks since it uses the basis of the attack model to determine where to insert key gates. The input pattern and corresponding key values can be analyzed during the insertion process of the obfuscation technique.

In 2017, TTLock was proposed that resisted all known attacks including SAT and sensitization attacks~\cite{ttlock}. TTLock would invert the response to a logic cone to protect the input pattern. The logic cone would be restored only if the correct key is provided. The small change to the functionality of the circuit would maximize the efforts needed for the SAT attacks. The generalized form of stripping away the functionality of logic cones and hiding it from attackers is known as stripped-functionality logic locking (SFLL). However, the design of TTLock didn't account for the cost of tamper-proof memory which could lead to high overhead in the re-synthesis process~\cite{atpg}~\cite{provably}. Another group automated the general process of TTLock to identify the parts of the design that needed to be modified in an efficient way. They used ATPG tools to develop a scalable and more efficient way of protecting these patterns from attackers. Overall, a 35\% improvement in overhead was achieved with the automated process. Later, a modified version of SFLL was proposed based on the hamming distance of the key. This was referred to as SFLL-hd~\cite{sfll-hd}. The hamming distance metric was used to determine which pattern to modify in the SFLL scheme. Depending on the type of attack, the hamming distance can be adjusted accordingly. In 2019, the idea of exploring high-level synthesis (HLS) with logic locking was proposed with SFLL-HLS~\cite{sfll-hls}. SFLL-HLS was proposed to improve the system-wide security of an IC. The design resulted in faster validation of design and higher levels of abstraction. The HLS implementation in this technique was used to identify the functional units and logic cones to be operated on with respect to SFLL. They observed low overhead and power results from their analysis. Most recently in 2020, LoPher was developed as another SAT resistant ofuscation technique~\cite{lopher}. LoPher uses a block cipher to produce the same behavior as a logic gate. The basic component for the block cipher is configurable and allows many logic permutation to occur which further increases the security of the system. 

Many forms and variations of SAT attacks have been created in order to show the weaknesses of various hardware obfuscation techniques. Algorithms have been developed for SAT competitions and the results can be used in a variety of applications including hardware obfuscation~\cite{host_sat}~\cite{lingeling}. These tools are used to evaluate the strength of logic locking techniques and can be used to bypass the security of integrated circuits. As a result, an anti-SAT unit was developed as a general solution to the SAT attack~\cite{antisat}. The anti-SAT block consists of a low overhead unit that can be added to any obfuscation technique to help counter the SAT attacks. The unit requires the key length for the locked circuit to increase as inputs to the anti-SAT block. The number of DIPs and input patterns that an adversary needs would grow exponentially due to this change. This would make the complexity of the SAT attack exponential instead of linear and therefore inefficient. The recent innovation in anti-SAT has inspired us to develop a technique that will be resistant to various SAT attacks. We designed constraints that should minimize the effects of a SAT algorithm.

\section{ProbLock}
ProbLock is based on filtering out nodes in a circuit to find the best location to insert key gates. ProbLock is a logic locking technique where the key gates are either XOR or XNOR gates and a key is used to unlock the circuit. We used four constraints to determine the best candidate nodes to insert our XOR or XNOR key gate; \textbf{longest path, non-critical path, low dependent nodes, and best probability nodes}. The first three constraints find the set of nodes that lie on the longest path, non-critical path, and have low dependent wires. The last constraint uses probability to find the set of nodes equal to the key length where we will insert the key gates. We chose the longest path and non-critical path constraint in order to avoid critical timing elements and to insert key gates on parts of the circuit that was being used the most. We chose the low dependent wires and probability constraint to determine locations where the output would be changed the most. This would make it harder for an attack to generate the golden circuit using an oracle based attack. Once we determine the location of the key nodes, we can insert key gates into the netlist and re-synthesize the circuit. In Equation~\ref{equ:constraints} the candidate nodes are determined from a function of all four constraints. $LP$ is the set of nodes on the longest paths while $NCP$ is the set of nodes on non-critical paths. $LD$ represents the set of low dependent nodes and $P$ is the set of probability nodes. 
\begin{equation}\label{equ:constraints}
    selectedNodes \subset{P} \subset{LD} \subset{NCP} \subset{LP}
\end{equation}
For our obfuscation technique, we decided to lock a set of combinational and sequential circuit netlists using the ISCAS '85 and ISCAS '89 circuit benchmarks. We obfuscated a total of 40 benchmarks using ProbLock. For some of the constraints, we had to use an unrolling technique described in~\cite{unroll} to accurately filter out nodes. This unrolling technique was only used in sequential circuits to simplify the concepts of flip flops and other sequential logic. The sequential logic can be replaced by the main stage and a $k$ number of sub-stages depending on the number of times unrolled. This results in a $k$-unrolled circuit that has the same functionality as the regular circuit. For this process, we generated a set of unrolled ISCAS '89 benchmarks which we used in some constraint algorithms. We unrolled these circuits once to prevent inaccuracies in constraints such as the longest path and non-critical path. 

\subsection{Longest Path Constraint}
\begin{figure}[h]
\centering
    \includegraphics[scale=.47]{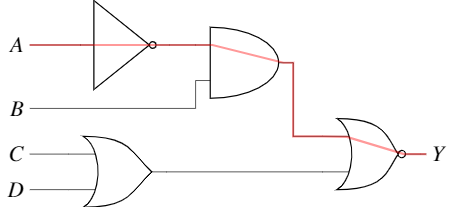}
\caption{Longest Path (in red)}
\label{fig:longest_path}
\end{figure}

The longest path constraint isolates a subset of nodes that lie on the longest paths in a circuit netlist. The subset of nodes is different for each circuit and is a function on the key length determined for each circuit. We represent the netlist of each benchmark as a directed acyclic graph (DAG) and perform the longest path analysis on each DAG. Each vertex in the DAG is a gate element from the netlist and each vector represents the wire connecting to the next gate element. Once the DAG is constructed for each benchmark, we calculated the longest paths of the DAG using a depth first search (DFS) technique. We then calculate the next longest path to generate a subset of nodes along the longest paths. Each unique node in the longest path gets added to a subset during each iteration until the size of the subset is bigger than two times the key length for that circuit. The structure of this theory is shown in Algorithm~\ref{alg:longest_path} which uses the DFS in Algorithm~\ref{alg:dfs}. Figure~\ref{fig:longest_path} shows the longest path for the circuit to be 3 since there are 3 gates between input $A$ and output $Y$. The next longest path would also be 3 from input $B$ to output $Y$. All of the nodes along both longest paths would be added to a subset of the longest path nodes. Once this subset of longest path nodes is determined, that subset gets used in the next filtering constraint. This subset can be adjusted to include more or fewer nodes depending on other filtering constraints. If more nodes are needed, this constraint is the first to be modified. 

We chose to use the longest path constraint in order to counter oracle guided attacks. Oracle guided attacks will query the IC with various inputs and observe the output. This gives the attacker information about how the circuit behaves and the adversary can use this information to determine the secret key. We want to insert key gates where most of the logic and activity occur in the circuit. An oracle guided attack will most likely pass data through the longest paths of a circuit so we want to protect these parts of the IC by inserting key gates on the longest path.
\begin{algorithm}
\SetAlgoLined
\caption{Get Longest Path}
\label{alg:longest_path}
\SetKwInOut{Input}{input}\SetKwInOut{Output}{output}
\Input{Circuit Graph and Key Length}
\Output{List of nodes on the longest path}
G$\leftarrow$ circuit graph\;
V$\leftarrow$ source vertex of G\;
overallNodes $\leftarrow$ []\;
\While{true}{
    allPaths $\leftarrow$ DFS(G,V)\;
    \For{p in allPaths}{
        \If{len(p) = maxLength}{
            maxPath $\leftarrow$ len(p)\;
        }
    }
    \For{p in maxPath}{
        \If{p not in overallNodes}{
            overallNodes.append(p)\;
        }
    }
    \If{len(overallNodes $>$ keyLength * 3)}{
        \algorithmicreturn{ overallNodes}\;
    }
}
\end{algorithm}
\begin{algorithm}
\SetAlgoLined
\caption{Depth First Search}
\label{alg:dfs}
\SetKwInOut{Input}{input}\SetKwInOut{Output}{output}
\Input{Circuit Graph G and Vertex Source V}
\Output{Longest path in Circuit}
mark V as visited\;
\For{all neighbors W of V}{
    \If{W is not visited}{
        DFS(G,W)\;
    }    
}
\end{algorithm}

\vspace*{-.5cm}

% \vspace*{-.5cm}
% \begin{figure}[h]
% \centering
% \begin{circuitikz}[scale = 0.7, transform shape]
%      \draw
%      (0,0) node[or port] (or1) {}
%      (-0.5,2) node[not port] (not1) {}
%      (2,1.72) node[and port] (and1) {}
%      (4,0.28) node[nor port] (nor1) {}
%      (not1.in 1) -- (-2,2) node[label={[label distance=-0.15cm]180:$A$}] {}
%      (and1.in 2) |- (-2,1) node[label={[label distance=-0.15cm]180:$B$}] {}
%      (or1.in 1) -- (-2,0.28) node[label={[label distance=-0.15cm]180:$C$}] {}
%      (or1.in 2) -- (-2,-0.28) node[label={[label distance=-0.15cm]180:$D$}] {}
%      (not1.out) |- (and1.in 1)
%      (and1.out) |- (nor1.in 1)
%      (or1.out) |- (nor1.in 2)
%      (nor1.out) |- (4.25,0.28) node[label={[label distance=-0.15cm]0:$Y$}] {}
%      ;
     
%      \draw[color=red, line width=1.5pt, opacity=0.4]
%      (-2,2) -- (0.88,2) -- (2, 1.72) -- (2.15,1.72) |- (nor1.in 1)
%      (nor1.in 1) -- (3,0.55) -- (3.9,0.28) -- (4.25,0.28);
%     \end{circuitikz}
% \caption{Longest Path (in red)}
% \label{fig:longest_path}
% \end{figure}

\subsection{Critical Path Constraint}
The critical path constraint is similar to the longest path; however, rather than considering logic depth, we look at timing information. This constraint is essential, as adding gates on the critical path could break the circuit functionality or change timing specifications. The nodes selected often overlapped with other constraints (e.g. the longest path was often the critical path), though oftentimes the critical path would involve gates with large fan out. Determining the critical path is largely technology-specific; different PDKs will have different timing information which can affect which paths are critical paths. We removed any nodes that were on the critical path from the set of nodes passed into this constraint. The resulting subset results in nodes that are on the longest, non-critical path. 

\begin{figure}[h]
\centering
    \includegraphics[scale=.47]{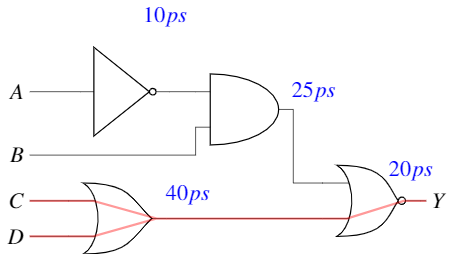}
\caption{Critical Paths (in red)}
\label{fig:critical_path}
\end{figure}

% \begin{figure}[h]
% \centering
% \begin{circuitikz}[scale = 0.7, transform shape]
%      \draw
%      (0,0) node[or port, label={[label distance=0.1cm]80:\textcolor{blue}{$40ps$}}] (or1) {}
%      (-0.5,2) node[not port, label={[label distance=0.2cm]80:\textcolor{blue}{$10ps$}}] (not1) {}
%      (2,1.72) node[and port, label={[label distance=0cm]80:\textcolor{blue}{$25ps$}}] (and1) {}
%      (4,0.28) node[nor port, label={[label distance=0.2cm]90:\textcolor{blue}{$20ps$}}] (nor1) {}
%      (not1.in 1) -- (-2,2) node[label={[label distance=-0.15cm]180:$A$}] {}
%      (and1.in 2) |- (-2,1) node[label={[label distance=-0.15cm]180:$B$}] {}
%      (or1.in 1) -- (-2,0.28) node[label={[label distance=-0.15cm]180:$C$}] {}
%      (or1.in 2) -- (-2,-0.28) node[label={[label distance=-0.15cm]180:$D$}] {}
%      (not1.out) |- (and1.in 1)
%      (and1.out) |- (nor1.in 1)
%      (or1.out) |- (nor1.in 2)
%      (nor1.out) |- (4.25,0.28) node[label={[label distance=-0.15cm]0:$Y$}] {}
%      ;
     
%      \draw[color=red, line width=1.5pt, opacity=0.4]
%      (-2,0.28) -- (-1,0.28) -- (0,0) -- (nor1.in 2) -- (3,0) -- (3.9,0.28) -- (4.25,0.28)
%      (-2,-0.28) -- (-1,-0.28) -- (0,0)
%      ;
%     \end{circuitikz}
% \caption{Critical Paths (in red)}
% \label{fig:critical_path}
% \end{figure}

\subsection{Low Wire Dependency Constraint}
The next constraint generates a subset of nodes that are connected to low dependent wires. The output wire of a gate is considered low dependent if the input wires to that gate have little influence on the value of output. This idea is modified from a technique called FANCI where suspicious wires can be detected in a Trojan infected design~\cite{fanci}. A functional truth table is created for each output wire of each gate in the circuit. The inputs of the truth table correspond to the inputs of the gate being analyzed. For each input column, the other columns are fixed and each row is tested with a 0 or 1 to determine the output. This results in two functions when setting the value to either 0 or 1. The boolean difference between these two functions results in a value between zero and one that can be further analyzed. The value for each input gets stored as a list for each output wire. We take the average value of the entire list to determine the dependency of an output wire. The algorithm logic is shown in Figure~\ref{alg:low_dependent_algo}. This analysis can determine if certain inputs are low dependent or if they rarely affect the corresponding logic. Low dependent wire are weak spots in the circuit so this constraint isolates those locations in order to improve the security. We insert key gates next to low dependent wires to fortify any weaknesses. The filtering process passes the subset of nodes to the final constraint.

\begin{algorithm}
\SetAlgoLined
\caption{Find Low Dependent Wires}
\label{alg:low_dependent_algo}
\SetKwInOut{Input}{input}\SetKwInOut{Output}{output}
\Input{Circuit Graph}
\Output{List of low dependent wires}
    $G$ $\leftarrow$ circuit graph\; 
    
    \ForEach{gates in G}{
        %$controlNodes$ $\leftarrow$ []\;
        \ForEach{output wire $w$}{
            $T$ $\leftarrow$ Truthtable($w$)\;
            $L$ $\leftarrow$ empty list of control values\;
            \ForEach{column c in $T$}{
                $count$ $\leftarrow$ 0\;
                \ForEach{row $r$ in $T$}{
                    $x_0$ $\leftarrow$ Value of $w$ when input value = 0\;
                    $x_1$ $\leftarrow$ Value of $w$ when input value = 1\;
                    \If{$x_0$ != $x_1$}{
                        $count$++\;
                    }
                }
                $L$.append\(\frac{count}{size(T)}\)\;
            }
            $avg$ $\leftarrow$ average($L$))\;
            \If{$avg$ $<$ 0.5}{
                $controlNodes$.append(gate)\;
            }
        }
    }
\end{algorithm}

\subsection{Biased Probabilities Constraint}
The probability constraint focuses on reducing the effectiveness of the SAT attacks. In a SAT attack, a distinguishing input (DI) is chosen and the attacker runs through various key values, eliminating any which yield an incorrect output. Thus, to reduce the effectiveness of a SAT attack, the number of wrong keys produced for a given DI must decrease. This can be done by bringing the probability of any given node being $1$ closer to $0.5$, since any node which is biased towards 0 or 1 will propagate through to the output nodes, making it easier for SAT attacks to eliminate key values. Since a two-input XOR/XNOR has an output probability of $0.5$, we can insert our key gates at nodes heavily biased towards 0 or 1 and "reset" the probability to $0.5$. 

The algorithm used to obtain the $N$ nodes with the most biased probabilities is shown in Algorithm~\ref{alg:prob_algo}. It is worth noting that while generating node probabilities for combinational circuits is trivial, sequential circuits pose a potential problem because of the D flip flops (DFFs). However, giving the DFF outputs a starting probability of 0.5 and propagating running a few iterations (three is sufficient) will asymptotically approach the correct probability for the DFF node.
\begin{algorithm}
\SetAlgoLined
\caption{Find Biased Nodes}
\label{alg:prob_algo}
\SetKwInOut{Input}{input}\SetKwInOut{Output}{output}
 \Input{Circuit Graph and Circuit Input Probabilities}
 \Output{List of $N$ most biased nodes}
 \For{$i\leftarrow 1$ \KwTo $N$}{
 DFF initial probability$\leftarrow$0.5\;
 \While{any probability unknown}{
  \ForEach{node with unknown probability}{
   \If{all node input probabilities known}{
    Compute node output probability\;
    }
   }
  }
  $Node\leftarrow max(abs(circuit probabilities-0.5))$\;
  Add $Node$ to output list\;
  Insert XOR/XNOR gate at $Node$ in Circuit Graph\;
 }
\end{algorithm}

\begin{figure}[htb!]
  \centering
  \begin{subfigure}[b]{0.45\columnwidth}
    \centering
        \includegraphics[scale=0.47]{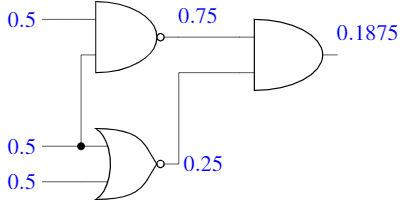}
    \caption{Pre-Insertion Probabilities}
    \label{fig:prob1}
  \end{subfigure}
  \hfill %%
  \begin{subfigure}[b]{0.5\columnwidth}
    \centering
        \includegraphics[scale=0.47]{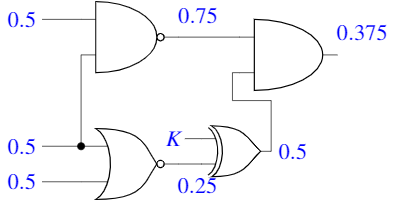}
    \caption{Post-Insertion Probabilities}
    \label{fig:prob2}
  \end{subfigure}
  \caption{Key Gate Insertion Probabilities}
  \label{fig:probability}
\end{figure}

An example of this is illustrated in~Figure~\ref{fig:probability}. Figure~\ref{fig:probability}~\subref{fig:prob1} shows a sample circuit with each node annotated with the probability of that node being a logic 1. The output shown is heavily biased toward logic 0, which makes it more susceptible to SAT attacks. Strategically adding a key gate, as shown in Figure~\ref{fig:probability}~\subref{fig:prob2} brings the output probability closer to 0.5, reducing the effectiveness of the SAT attack.

\section{Implementations}
To obfuscate the benchmarks, we created a python script that implements this algorithm. The script takes in a benchmark netlist in Verilog format and returns an obfuscated netlist in the same format. The obfuscated netlist included the key gates inserted as well as the key defined to unlock the circuit. We created a function to parse each netlist for information. The information was organized into lists of inputs, outputs, and gate types. We used this information to determine the key size relative to the number of gates in a netlist. We also created functions for each constraint in our algorithm. A set of overall nodes was passed through each function and then narrowed down to a set of best nodes for key gate insertion. Another function was created to insert key gates from a data structure into a new netlist. We specified the key inputs, key gates, and the key value in the header of the new netlist for development purposes. Throughout the development process, we ran tests to verify the intention of our script and to make sure each new netlist was correct.

 We used Synopsys Design Compiler to synthesize and view the netlists before and after obfuscation~\cite{DC}. The Synopsys tool allowed us to see the gate level representation of each benchmark during the analysis process. We were able to see the location of the logic components as well as the inserted key gate after the obfuscation process occurred. We also used the Design Compiler for critical path analysis in our second constraint. The tool allowed for timing analysis between different logic components of the netlist. We used this to calculate the critical paths for our constraint and removed any nodes that lie on this critical path. We integrated the results from Synopsys by passing it through a textfile that gets parsed in the main script.
 
 \section{Experimental Results}
During the development process of ProbLock, we analyzed the correlation and relationship between constraints. We documented this relationship to show how each constraint impacted the overall strength of the technique. We chose two constraints based on path elements and two constraints based on nodes and wires. Due to this design, we were able to analyze the correlation between constraints and adjust the strength of the filtering process based on this analysis. Overall, we wanted the correlation between constraints to be large enough to remove any nodes that didn't belong in both sets. This would allow the filtering process from each constraint to generate a subset of nodes each time until only the best candidate nodes remain to be inserted. The strength of the correlation varies between benchmarks because of the shape and functionality of each circuit. Each subsequent constraint filtered out a set of nodes based on the relationship between the constraint and the overall set of nodes. Table~\ref{tab:correlation} shows the experimental correlations for ISCAS '85 and '89 benchmarks. We only show some of the results in the table as all 40 benchmarks are not included. The longest path (LP) length and critical path (CP) count are based on the path constraints. The rest of the categories including non-critical path (NCP), low depending wires (LD), and biased probabilities (Prob) are based on nodes. 

For each constraint, a smaller set of nodes is generated until final set is determined which corresponds to the location of inserted key gates. In the ISCAS '85 benchmark suite, the correlation between nodes on the longest path, and the original set of nodes was 36\% on average. Between nodes on the non-critical path and nodes on the longest path, the correlation was about 63\% on average. The correlation between low dependent nodes and nodes on the non-critical path was about 73\% and the final correlation between biased probabilities and low dependant nodes was about 65\%. For the ISCAS '89 benchmark suite, the correlation between the longest path and overall nodes is 27\%. The correlation between the critical path and the longest path is 84\%. The correlation between low dependent nodes and non-critical path is 85\% and the final correlation between biased probabilities and low dependent nodes is 45\%. The numbers that we analyzed were ideal for the filtering process. Enough nodes were removed with each subset until the final set of best candidates were discovered. For the final biased probabilities constraint, the final set of nodes was equal to the size of the key. For the other constraints, we adjusted the filtering threshold accordingly. Depending on the situation, the strength of the constraints can be adjusted which allows flexibility in our algorithm. After re synthesising the obfuscated netlists, we used Synopsys to verify the behavior of the locked circuits~\cite{DC}. The critical path of the netlists remained the same and the timing analysis remained consistant for all benchmarks. We also used Synopsys to verify that the overhead of netlist was no greater than 10\%. 

\begin{table*}[]
\small
\caption{ISCAS '85 \& '89 Constraint Correlation}\label{tab:correlation}
\begin{tabular}{|l|r|r|r|r|r|r|r|r|}
\hline
\textbf{ISCAS 85} & \multicolumn{1}{l|}{\textbf{Key Size}} & \multicolumn{1}{l|}{\textbf{LP Length}} & \multicolumn{1}{l|}{\textbf{CP Count}} & \multicolumn{1}{l|}{\textbf{Total Nodes}} & \multicolumn{1}{l|}{\textbf{LP Subset}} & \multicolumn{1}{l|}{\textbf{NCP Subset}} & \multicolumn{1}{l|}{\textbf{LD Subset}} & \multicolumn{1}{l|}{\textbf{Prob Subset}} \\ \hline
c432 & 16 & 18 & 7 & 160 & 88 & 60 & 33 & 16 \\ \hline
c499 & 16 & 12 & 32 & 202 & 186 & 104 & 99 & 16 \\ \hline
c1355 & 32 & 25 & 32 & 546 & 485 & 253 & 53 & 32 \\ \hline
c1908 & 64 & 39 & 25 & 880 & 205 & 145 & 129 & 64 \\ \hline
c2670 & 64 & 31 & 100 & 1269 & 217 & 200 & 75 & 64 \\ \hline
c3540 & 128 & 42 & 22 & 1669 & 260 & 173 & 151 & 128 \\ \hline
c5315 & 128 & 47 & 100 & 2307 & 411 & 226 & 180 & 128 \\ \hline
c7552 & 256 & 35 & 100 & 3513 & 532 & 341 & 278 & 256 \\ \hline
\textbf{ISCAS 89} & \multicolumn{1}{l|}{\textbf{Key Size}} & \multicolumn{1}{l|}{\textbf{LP Length}} & \multicolumn{1}{l|}{\textbf{CP Count}} & \multicolumn{1}{l|}{\textbf{Total Nodes}} & \multicolumn{1}{l|}{\textbf{LP Subset}} & \multicolumn{1}{l|}{\textbf{NCP Subset}} & \multicolumn{1}{l|}{\textbf{LD Subset}} & \multicolumn{1}{l|}{\textbf{Prob Subset}} \\ \hline
s298 & 8 & 10 & 6 & 75 & 26 & 26 & 18 & 8 \\ \hline
s344 & 8 & 21 & 11 & 101 & 21 & 19 & 19 & 8 \\ \hline
s382 & 8 & 10 & 6 & 99 & 29 & 29 & 21 & 8 \\ \hline
s386 & 8 & 12 & 7 & 118 & 49 & 32 & 24 & 8 \\ \hline
s400 & 8 & 10 & 6 & 106 & 30 & 30 & 22 & 8 \\ \hline
s444 & 8 & 12 & 6 & 119 & 38 & 38 & 30 & 8 \\ \hline
s526 & 8 & 10 & 6 & 141 & 26 & 26 & 18 & 8 \\ \hline
s641 & 8 & 75 & 24 & 107 & 80 & 53 & 51 & 8 \\ \hline
s713 & 8 & 75 & 23 & 139 & 84 & 61 & 56 & 8 \\ \hline
s838 & 16 & 18 & 1 & 288 & 45 & 29 & 26 & 16 \\ \hline
s1238a  & 32  & 23 & 14  & 428   & 132 & 76  & 64  & 32  \\ \hline
s1488 & 32 & 18 & 19 & 550 & 89 & 60 & 48 & 32 \\ \hline
s5378a  & 64  & 15 & 46  & 1004  & 134 & 96  & 92  & 64  \\ \hline
s9234a & 128 & 19 & 37 & 2027 & 264 & 261 & 213 & 128 \\ \hline
s13207a & 256 & 28 & 100 & 2573 & 573 & 521 & 482 & 256 \\ \hline
s15850a & 256 & 22 & 100 & 3448 & 553 & 544 & 506 & 256 \\ \hline
s38584 & 256 & 15 & 100 & 11448 & 717 & 716 & 571 & 256 \\ \hline
\end{tabular}
\end{table*}

\section{Conclusions and Future Work}
We propose ProbLock, a probability-based logic locking technique that uses a filtering process to determine the location of inserted key gates. ProbLock uses four constraints to narrow the set of nodes in a netlist to be used for insertion. We obfuscated $40$ different sequential and combinational benchmarks from the ISCAS '85 and ISCAS '89 suite. After obfuscating the circuits, we analyzed the correlation between constraints and implemented the capability to adjust these constraints depending on the situation. In the future, we intend to test the obfuscated benchmarks against known attacks and compare it to other logic locking techniques. We will implement logic locking attacks such as SAT attacks and sensitization attacks. Each attack will be executed against benchmarks obfuscated with ProbLock. We will then run the same attacks on locking schemes such as SLL~\cite{onimproving}, logic cone locking~\cite{logiccone}, and SARLock~\cite{sarlock}. We will evaluate how well each benchmark performs by measuring overhead of the obfuscation technique, complexity of the technique, and execution time of the attack. After running each attack scheme, we can compare and evaluate the true strength of ProbLock compared to other published logic locking techniques. We also plan on strengthening ProbLock against SAT attacks by integrating SAT resistant logic near the key gate locations. This would increase overhead, but we would also try to optimize this in our experiment.


\begin{thebibliography}{10}

\bibitem{tehranipoor2018low}
Jake Mellor, Allen Shelton, Michael Yue, and Fatemeh Tehranipoor.
\newblock Attacks on logic locking obfuscation techniques.
\newblock In {\em 2021 IEEE International Conference on Consumer Electronics
  (ICCE)}, pages 1--6, 2021.

\bibitem{book}
D.~{Forte}, S.~{Bhunia}, and M.~{Tehranipoor}.
\newblock Hardware protection through obfuscation.
\newblock 2017.

\bibitem{logiccone}
Y.~{Lee} and N.~A. {Touba}.
\newblock Improving logic obfuscation via logic cone analysis.
\newblock {\em 16th Latin-American Test Symposium (LATS)}, pages 1--6, 2015.

\bibitem{onimproving}
M.~{Yasin}, J.~J.~V. {Rajendran}, O.~{Sinanoglu}, and R.~{Karri}.
\newblock On improving the security of logic locking.
\newblock {\em IEEE Transactions on Computer-Aided Design of Integrated
  Circuits and Systems}, 35(9):1411--1424, 2016.

\bibitem{overview}
F.~{Tehranipoor}, W.~{Yan}, and J.~A. {Chandy}.
\newblock Development and evaluation of hardware obfuscation benchmarks.
\newblock {\em J Hardware System Security 2}, pages 142--161, 2018.

\bibitem{sarlock}
M.~{Yasin}, B.~{Mazumdar}, J.~J.~V. {Rajendran}, and O.~{Sinanoglu}.
\newblock Sarlock: Sat attack resistant logic locking.
\newblock {\em IEEE International Symposium on Hardware Oriented Security and
  Trust (HOST)}, pages 236--241, 2016.

\bibitem{tehranipoor2019deep}
Fatemeh Tehranipoor, Nima Karimian, Mehran Mozaffari~Kermani, and Hamid
  Mahmoodi.
\newblock Deep rnn-oriented paradigm shift through bocanet: Broken obfuscated
  circuit attack.
\newblock In {\em Proceedings of the 2019 on Great Lakes Symposium on VLSI},
  pages 335--338, 2019.

\bibitem{iscas85}
F.~{Brglez} and H.~{Fujiwara}.
\newblock A neutral netlist of 10 combinational benchmark circuits and a target
  translator in fortan.
\newblock {\em Proc. of the International Symposium on Circuits and Systems},
  pages 663--698, 1985.

\bibitem{iscas89}
F.~{Brglez}, D.~{Bryan}, and K.~{Kozminski}.
\newblock Combinational profiles of sequential benchmark circuits.
\newblock {\em Proc. of the International Symposium of Circuits and Systems},
  pages 1929--1934, 1989.

\bibitem{ttlock}
M.~{Yasin} and O.~{Sinanoglu} B.~{Mazumdar}, J. J. V.~{Rajendran}.
\newblock Ttlock: Tenacious and traceless logic locking.
\newblock {\em 2017 IEEE International Symposium on Hardware Oriented Security
  and Trust (HOST)}, pages 166--166, 2017.

\bibitem{atpg}
A.~{Sengupta}, M.~{Nabeel}, M.~{Yasin}, and O.~{Sinanoglu}.
\newblock Atpg-based cost-effective, secure logic locking.
\newblock {\em 2018 IEEE 36th VLSI Test Symposium (VTS)}, pages 1--6, 2018.

\bibitem{provably}
M.~{Yasin}, A.~{Sengupta}, and J.~J. V. {Rajendran} O.~{Sinanoglu} M.~{Thari
  Nabeel}, M.{Ashraf}.
\newblock Provably-secure logic locking: From theory to practice.
\newblock {\em CCS '17: Proceedings of the 2017 ACM SIGSAC Conference on
  Computer and Communications Security}, pages 1601--1618, 2020.

\bibitem{sfll-hd}
F.~{Yang}, M.~{Tang}, and O.~{Sinanoglu}.
\newblock Stripped functionality logic locking with hamming distance-based
  restore unit (sfll-hd) - 2013 unlocked.
\newblock {\em IEEE Transactions on Information Forensics and Security},
  14(10):2778--2786, 2019.

\bibitem{sfll-hls}
C.~{Zhao} J. J. V.~{Rajendran} M.~{Yasin}.
\newblock Sfll-hls: Stripped-functionality logic locking meets high-level
  synthesis.
\newblock {\em 2019 IEEE/ACM International Conference on Computer-Aided Design
  (ICCAD)}, pages 1--4, 2019.

\bibitem{lopher}
A.~{Saha}, S.~{Saha}, and B.~B.~{Bhattacharya} S.~{Chowdhury},
  D.~{Mukhopadhyay}.
\newblock Lopher: Sat-hardened logic embedding on block ciphers.
\newblock {\em 2020 57th ACM/IEEE Design Automation Conference (DAC)}, pages
  1--6, 2020.

\bibitem{host_sat}
P.~{Subramanyan}, S.~{Ray}, and S.~{Malik}.
\newblock Evaluating the security of logic encryption algorithms.
\newblock In {\em 2015 IEEE International Symposium on Hardware Oriented
  Security and Trust (HOST)}, pages 137--143, 2015.

\bibitem{lingeling}
A.~{Biere}.
\newblock {Splatz, Lingeling, Plingeling, Treengeling, YalSAT Entering the SAT
  Competition 2016}.
\newblock In Tom\'{a}\v{s} Balyo, Marijn Heule, and Matti J{\"a}rvisalo,
  editors, {\em Proc.~of {SAT Competition} 2016 -- Solver and Benchmark
  Descriptions}, volume B-2016-1 of {\em Department of Computer Science Series
  of Publications B}, pages 44--45. University of Helsinki, 2016.

\bibitem{antisat}
Y.~{Xie} and A.~{Srivastava}.
\newblock Anti-sat: Mitigating sat attack on logic locking.
\newblock {\em IEEE Transactions on Computer-Aided Design of Integrated
  Circuits and Systems}, 38(22):199--207, 2019.

\bibitem{unroll}
N.~{Miskov-Zivanov} and D.~{Marculescu}.
\newblock "modeling and optimization for soft-error reliability of sequential
  circuits.
\newblock {\em IEEE Transactions on Computer-Aided Design of Integrated
  Circuits and Systems}, 27(5):803--816, 2008.

\bibitem{fanci}
A.~{Waksman}, M.~{Suozzo}, and S.~{Sethumadhavan}.
\newblock Fanci: Identification of stealthy malicious logic using boolean
  functional analysis.
\newblock {\em Proceedings of the 2013 ACM SIGSAC Conference on Computer \&
  Communications Security}, pages 697--708, 2013.

\bibitem{DC}
Design compiler graphical.
\newblock Synopsys, 2018.

\end{thebibliography}
\end{document}